\documentclass[final,1p,times]{elsarticle} 
\usepackage{graphicx} 
\usepackage{amssymb} 
\usepackage{amsthm} 
\usepackage{lineno} 

\journal{Nuclear Physics A} 
\begin{document} 

\begin{frontmatter} 


\title{Baryon Stopping in Au+Au and p+p collisions at 62 and 200 GeV}

\author{Hans Hjersing Dalsgaard for the BRAHMS Collaboration}

\address[a]{Niels Bohr Institute, 
University of Copenhagen,
Blegdamsvej 17, 2100 Ø, Denmark}

\begin{abstract} 
BRAHMS has measured rapidity density distributions of protons and
antiprotons in 
both p+p and Au+Au collisions at 62 GeV and 200 GeV. From these
distributions the 
yields of so-called `net-protons', that is the difference between the
proton and antiproton yields, 
can be determined.  The rapidity dependence of the net-proton yields
from  peripheral Au+Au collisions is found to have a similar behaviour
to that found for the p+p results, while a quite different rapidity
dependence  is found for central Au+Au collisions.  The net-proton
distributions can be used together with model calculations to find the
net-baryon yields as a function of rapidity, thus yielding information
on the average  rapidity loss of beam particles, the baryon
transport properties  of the medium, and the amount of `stopping'
in these collisions. 

\end{abstract} 

\end{frontmatter} 



\section{p+p and peripheral Au+Au}
\begin{figure}[ht]

\begin{minipage}{0.5\linewidth}
  \centering
  \includegraphics[width=\textwidth]{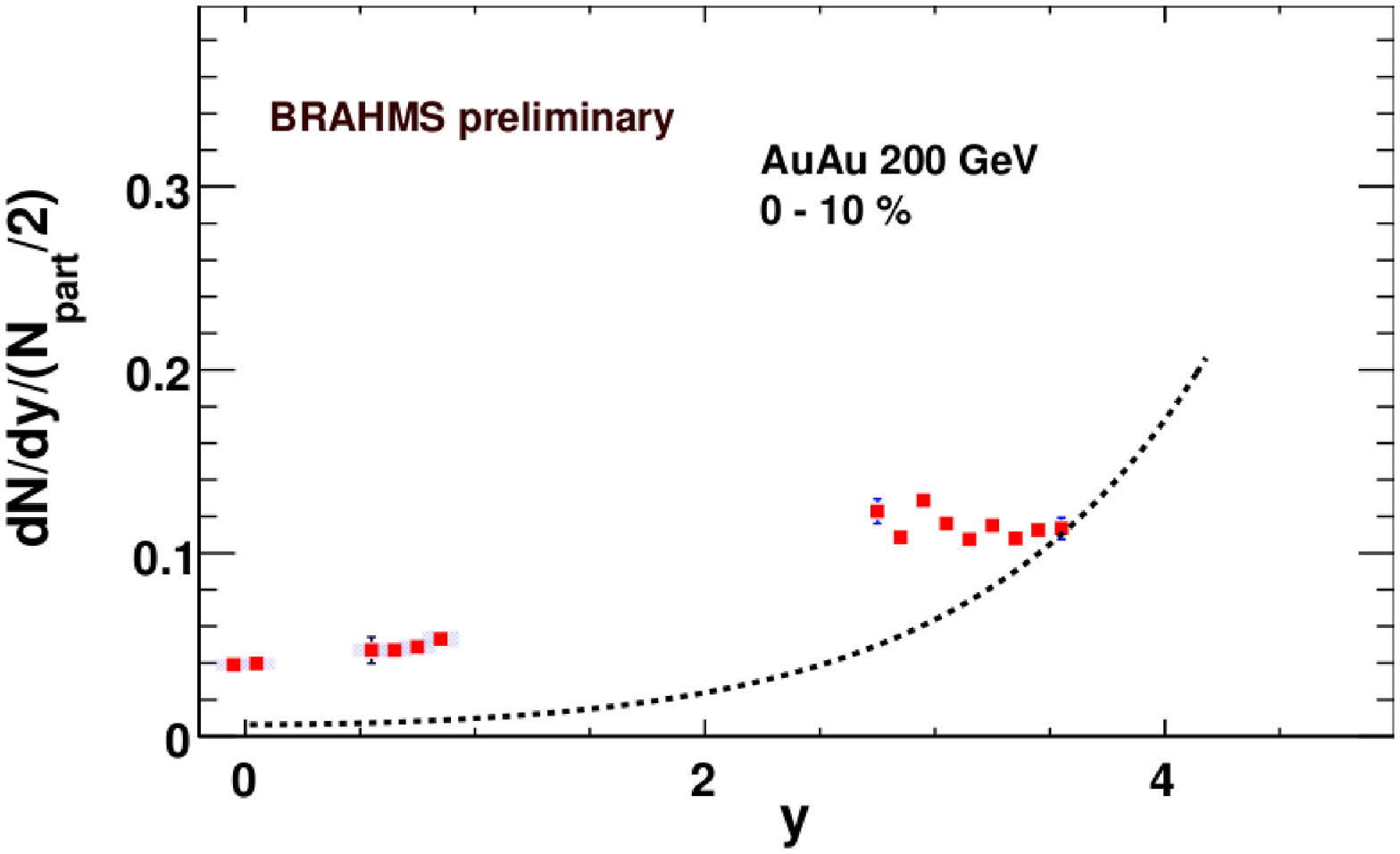}
  
  \end{minipage}%
  \begin{minipage}{0.5\linewidth}
  \centering
  \includegraphics[width=\textwidth]{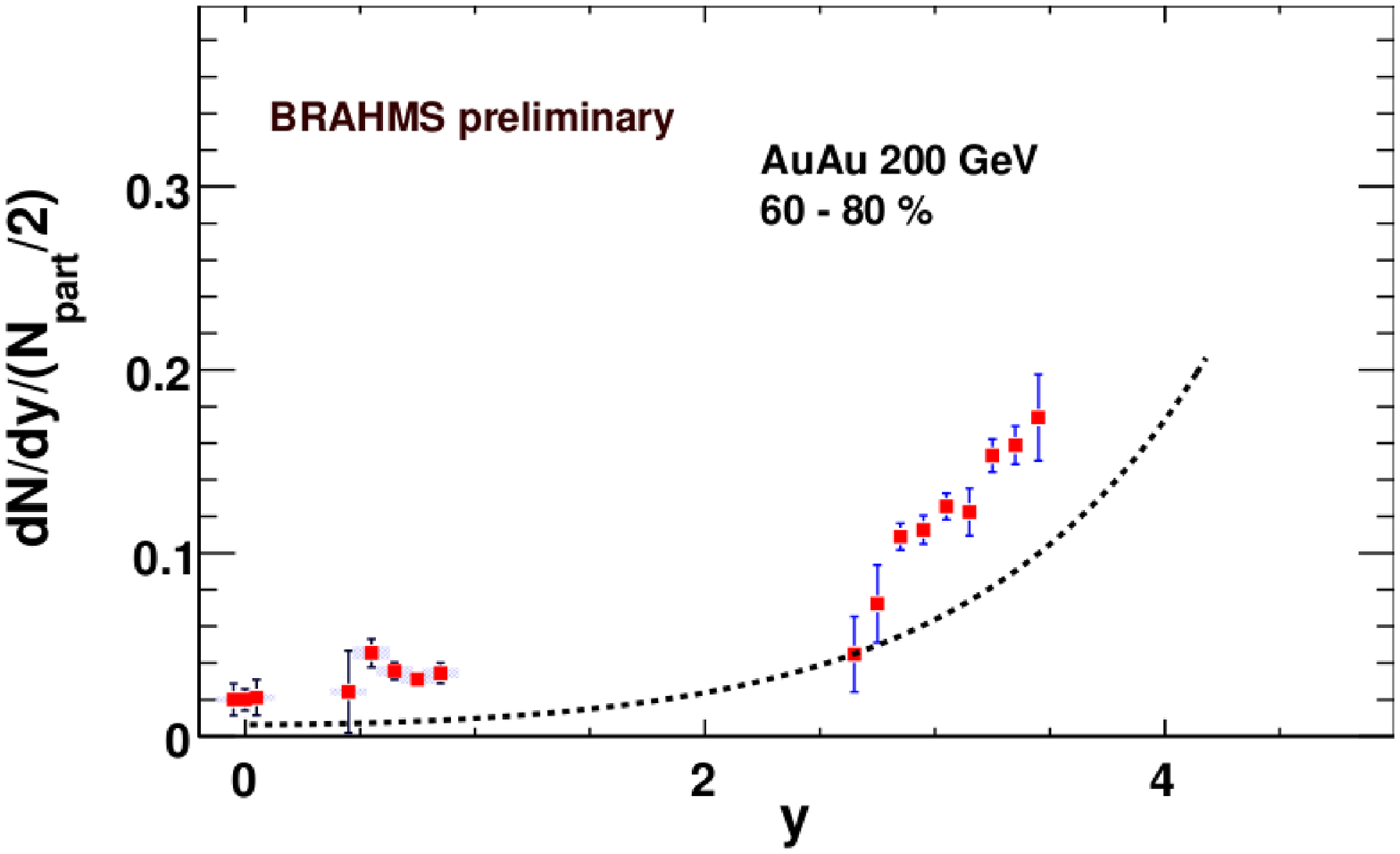}

  \end{minipage}
  \caption[]{Net-proton distributions from Au+Au collisions compared to the
    scaling observed for p+p collisions \cite{videbaek_QM09}. Left
    panel: Central collisions. Right panel: Peripheral collisions.}
\label{fig:scaling}
\end{figure}

In p+p collisions we expect that 
$\frac{dN}{dy'}$ where $y' = y-y_{b}$ ($y_b$ is the beam rapidity)
should follow an exponential 
in $y'$ and this behaviour is confirmed by BRAHMS p+p data
\cite{videbaek_QM09}. 
The right panel of Fig. \ref{fig:scaling} shows $\frac{dN}{dy'}$ from
peripheral 200 GeV Au+Au collisions scaled by $N_{part}$ ($N_{part}$
is the number of participants) overlaid with the exponential 
curve found for p+p collisions. It is seen that the two systems show
quantitatively 
similar dependence. This confirms that some aspects of p+p
collisions and peripheral Au+Au are very much alike. As a reference, the
left panel of Fig. 1 shows $\frac{dN}{dy'}$ for central Au+Au
collisions overlaid with the p+p scaling curve. It is evident that central
Au+Au and p+p collisions do not follow the same type of scaling. This
indicates that there are more collective stopping mechanisms in play
for central collisions.

\section{Baryon stopping}
\begin{figure}[ht]

\begin{minipage}{0.5\linewidth}
  \centering
  \includegraphics[width=\textwidth]{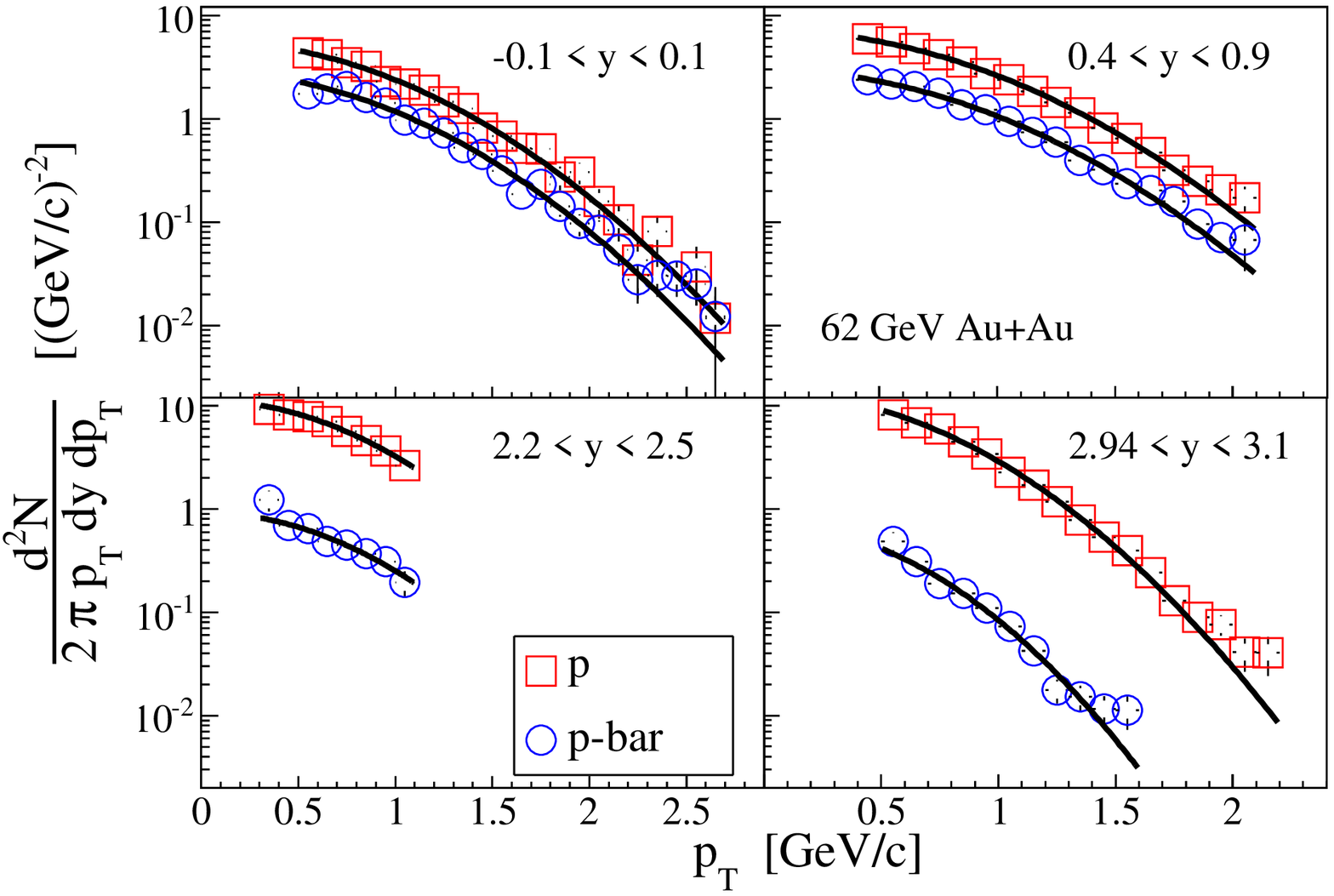}
  
  \end{minipage}%
  \begin{minipage}{0.5\linewidth}
  \centering
  \includegraphics[width=\textwidth]{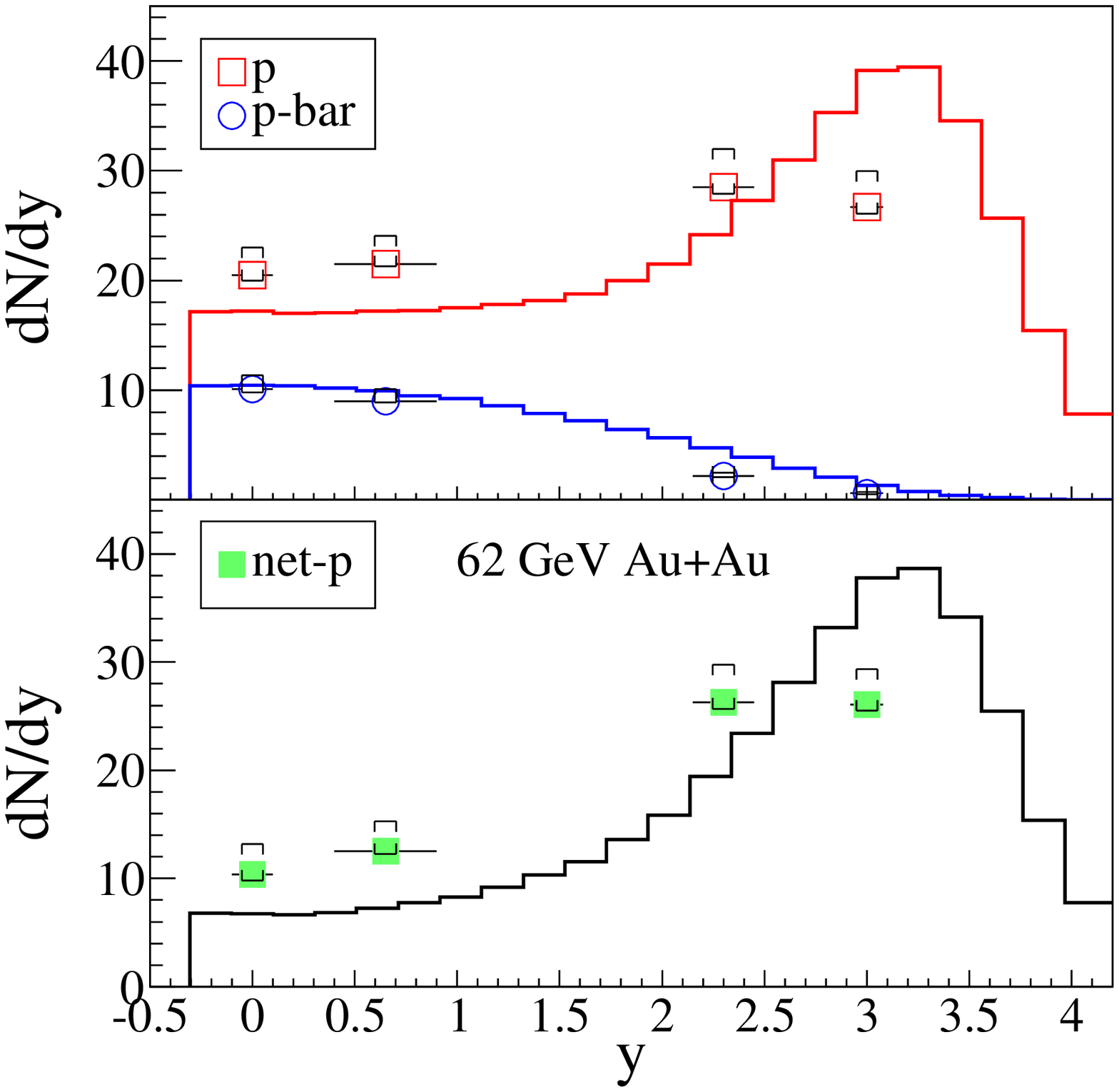}

  \end{minipage}
  \caption[]{Spectra and yields of identified protons and antiprotons and the
  resulting net-protons.}
\label{fig:yields}
\end{figure}

BRAHMS \cite{BRAHMSNIM} has measured the stopping in Au+Au
collisions at $\sqrt{s_{NN}}=200$ GeV \cite{BRAHMS}. Results from
$\sqrt{s_{NN}}=62.4$ GeV Au+Au collisions can be used to expand the
understanding of the stopping in the `energy gap' between the SPS top
energy of $\sqrt{s_{NN}}=17$ GeV and the RHIC top energy of 200 GeV. The left
panel of Fig. \ref{fig:yields} shows the proton and antiproton
spectra in four rapidity intervals. Corrections have been applied to
the data for 
geometrical acceptance, efficiency and detector effects such as
multiple scatterings. The right panel of Fig. \ref{fig:yields} shows
the extrapolated yields versus rapidity. The extrapolation was done
using a fit function of the form $f(p_T) \propto
\exp(-p_{T}^2/2\sigma^2)$. 
The
net-proton yields are also shown in the bottom panel of Fig. \ref{fig:yields}. Also
included in the figure are comparisons to HIJING/$B \bar{B}$  \cite{HIJING}. It
is seen that HIJING reproduces the anti-protons well but deviates from
the proton yields. This indicates that the baryon transport
description in HIJING/$B \bar{B}$ underpredicts stopping in central Au+Au collisions and is not sufficient to describe the data.

To quantify the stopping we use the average rapidity loss defined as \cite{VidebaekHansen}:
\begin{equation}
\delta y =  y_{b} -\frac{2}{N_{part}} \int_{0}^{y_{b}}
y\frac{dN_{B-\bar{B}}}{dy} dy 
\label{eq:AvRapLoss}
\end{equation}
Here, 
$\frac{dN_{B-\bar{B}}}{dy}$ is the number of net-baryons and $y_{b} = 4.2$ for $\sqrt{s_{NN}}=62.4$ GeV. Since BRAHMS does not 
measure neutrons or $\Lambda$'s we 
must make a conversion from net-protons based on simulations and data
from other experiments. For details of this procedure see \cite{BRAHMS62}
. The conversion used here is $\frac{dN_{B-\bar{B}}}{dy} = (2 \pm
0.1)\cdot \frac{dN_{p-\bar{p}}}{dy}$ at mid-rapidity and $\frac{dN_{B-\bar{B}}}{dy} = (2.1 \pm 0.1)\cdot \frac{dN_{p-\bar{p}}}{dy}$ at
forward rapidities (the larger correction at forward rapidities is due
to a small increase in the $n/p$ ratio from HIJING/$B \bar{B}$).

\begin{figure}[ht]
  \centering
  \includegraphics[width=0.6\linewidth]{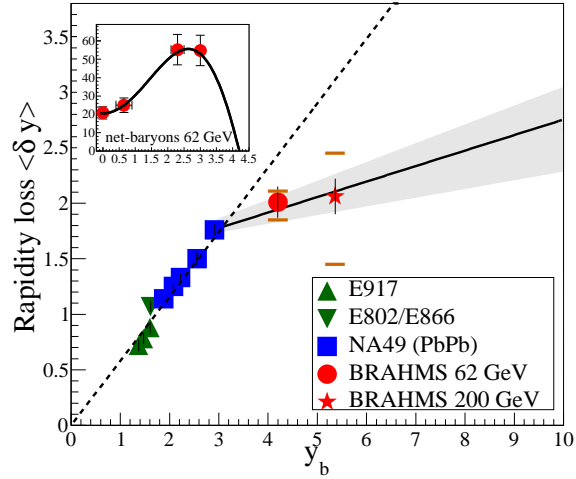}
  
\caption[width=0.6\linewidth]{Rapidity losses from  AGS
\cite{AGSE917, AGSE802, AGSE877}, SPS \cite{SPSNA49,NA49_prelim} and
RHIC \cite{BRAHMS}. The rapidity seems to saturate above SPS energies.}
\label{fig:raplosses}
\end{figure}

To calculate the rapidity loss we fit the resulting net-baryon
distribution with a third degree polynomial in $y^2$. This fit is shown as
the inset in Fig. \ref{fig:raplosses}.
The rapidity loss for $\sqrt{s_{NN}}=62.4$ GeV is measured to be
(stat. + syst. error): 
$$\delta y = 2.01 \pm 0.14 \pm 0.12 $$

Figure \ref{fig:raplosses} shows rapidity losses from AGS
\cite{AGSE917, AGSE802, AGSE877}, SPS \cite{SPSNA49,NA49_prelim}, and
RHIC \cite{BRAHMS}. The new $\sqrt{s_{NN}}=62.4$ GeV data from BRAHMS are seen to
establish that the apparent saturation of the rapidity losses sets in
already around the top SPS energy. 
\section{Limiting Fragmentation}
\begin{figure}[htb]

\begin{minipage}{0.5\linewidth}
  \centering
  \includegraphics[width=\textwidth]{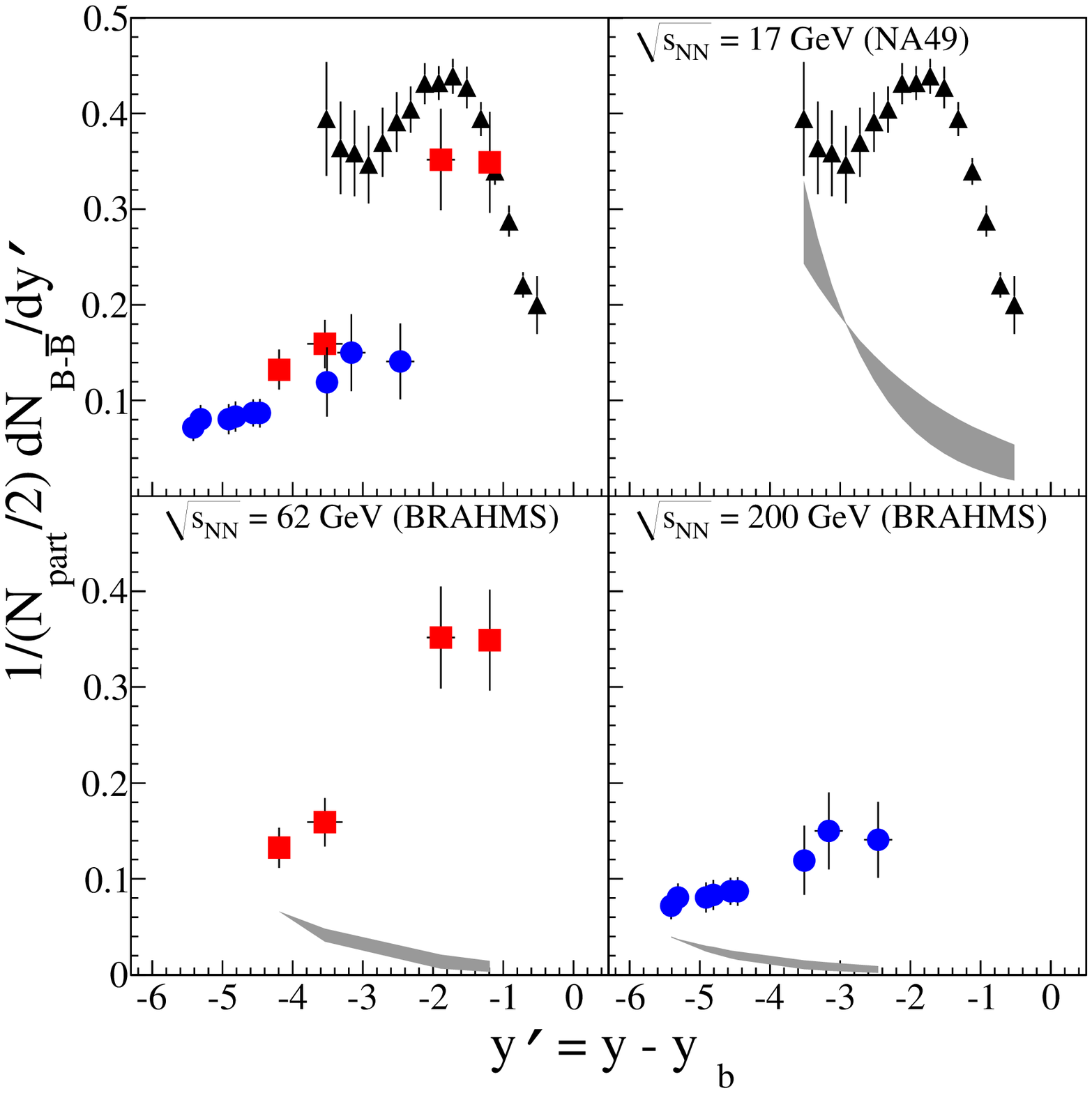}
  
  \end{minipage}%
  \begin{minipage}{0.5\linewidth}
  \centering
  \includegraphics[width=\textwidth]{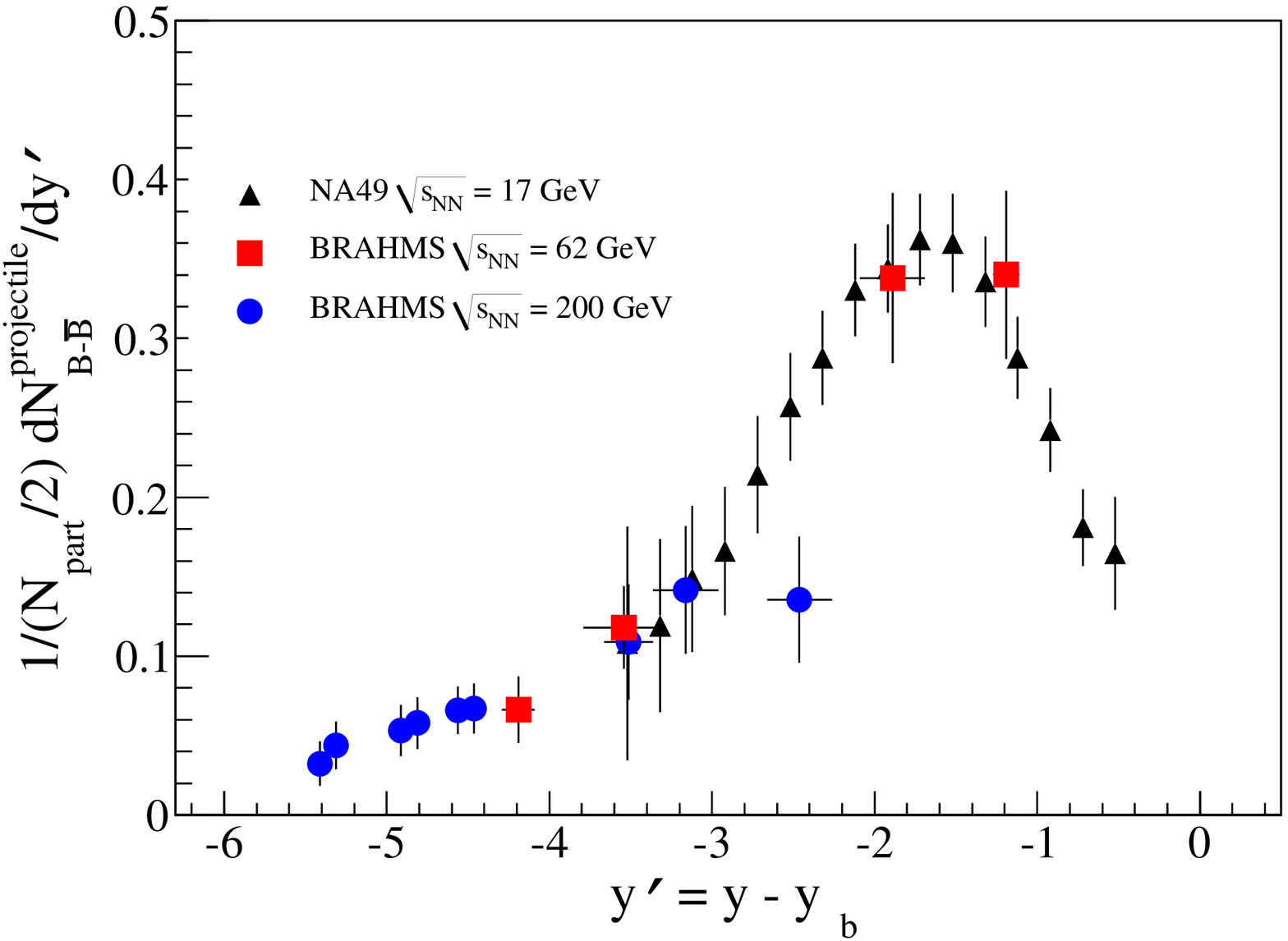}
  \end{minipage}

\caption[]{Left panel: $dN/dy'$ distributions from SPS
  \cite{SPSNA49,NA49_prelim} and RHIC \cite{BRAHMS} and their
  `target' distributions (grey bands). Right panel: The resulting `limiting
  fragmentation' distribution for SPS and RHIC data.}
\label{fig:lim_frag_comparison}
\end{figure}
Since there seems to be a linear increase of the average rapidity loss
from the SPS top energy to
the RHIC top energy we have studied if there exists some scaling of
the yields. The left panel of Fig. \ref{fig:lim_frag_comparison}
shows the yields from 
SPS and RHIC plotted versus $y'$ and it is easily seen that
there is no obvious universal behaviour.  

The idea is now to consider the yields in a `limiting fragmentation'
picture. We will do this by considering only one side of the collision
which we denote the `projectile' side of the collision inspired by
fixed target experiments. The challenge is now to remove the `target'
side of the distributions. We use two different estimates to set
limits for the `target' contribution: (1) a simple exponential form
$\exp(-y')$~\cite{BuszaGoldhaber84} and (2) a gluon 
junction motivated form $\exp(-y'/2)$~\cite{Kopeliovich:1988qm}. The
resulting estimates for the contributions from the `target' are shown as the grey bands in the
left panel of Fig. \ref{fig:lim_frag_comparison} together with the measured $dN/dy'$ distributions
from SPS and RHIC.

The right panel of Fig. \ref{fig:lim_frag_comparison} shows the
resulting `projectile' distributions from SPS and RHIC and it seen
that now we have a scaling behaviour between SPS and up to RHIC 62
GeV similar to limiting fragmentation. The stopping pattern in central
Au+Au collisions at 200 GeV shows some deviation from the trend which
suggests an energy dependence of the stopping mechanism.

\section{Conclusions}
BRAHMS has measured the rapidity loss in Au+Au collisions at
$\sqrt{s_{NN}}=62.4$ GeV which bridges the gap between the SPS top
energy and the RHIC top energy. The rapidity losses seem to saturate
from the SPS top energy and the saturating behaviour is confirmed by
the $\sqrt{s_{NN}}=62.4$ GeV data. Furthermore we have established a
limiting fragmentation kind of scaling in $dN/dy'$ distributions from
SPS to RHIC.

In these proceedings we have also 
demonstrated the similarity
between peripheral Au+Au collisions and p+p collisions using new BRAHMS data.


\end{document}